\documentclass{article}
\title{\bf Symmetric Vertex Models on \\
Planar Random Graphs
}

\author{ {\it D.A. Johnston}\\
         Dept. of Mathematics\\
         Heriot-Watt University\\
         Riccarton\\
         Edinburgh, EH14 4AS, Scotland
         }
\begin{document}
  \maketitle
                      {\Large
                      \begin{abstract}
%
We solve a 4-(bond)-vertex model on an ensemble of  3-regular ($\Phi^3$)
planar random graphs, which has the effect of coupling the vertex model
to 2D quantum gravity. 
The method of solution, by mapping onto an Ising model in field, 
is inspired by the 
solution by Wu et.al. of the regular lattice equivalent --
a symmetric 8-vertex model on the honeycomb lattice, and also applies
to higher valency bond vertex 
models on random graphs when the vertex weights
depend only on bond numbers and not cyclic ordering
(the so-called symmetric vertex models).

The relations
between the vertex weights and Ising model parameters in the 4-vertex model
on $\Phi^3$ graphs
turn out to be identical to those of the honeycomb lattice model,
as is the form of the 
equation of the Ising critical locus for the vertex weights.
A symmetry of the partition function under transformations of the
vertex weights, 
which is fundamental to the solution in both cases, can be understood in the
random graph case as a change
of integration variable in the matrix integral
used to define the model.

Finally, we note that vertex models, such as that discussed in this paper, may have a role
to play in the discretisation of Lorentzian metric quantum gravity in two dimensions.
%
                        \end{abstract} }
%
  \thispagestyle{empty}
%
%
  \newpage
%
                  \pagenumbering{arabic}

\section{Introduction}

Random matrix models have proved remarkably successful in investigating
the critical behaviour of various sorts of spin models (Ising \cite{bk,bk2},
Potts \cite{kaz}, $O(N)$ \cite{various} $\ldots$) living 
on ensembles of planar random graphs. This, in effect, couples the models
to $2D$ quantum gravity which originally motivated their study in the context
of string theory and random surfaces \cite{pol}.
From the purely statistical mechanical point of view one is looking 
at the critical behaviour 
of an (annealed) ensemble of random graphs decorated by the spins. The 
connection with gravity comes from the fact that the graphs
can adapt their connectivity in response to the spin configuration, which in
turn is influenced by the connectivity of the graph on which it lives. This 
captures the essential feature of the back reaction of the matter on the geometry
that characterises gravity in a discrete form.

The graphs of interest appear as planar Feynman diagrams in the 
perturbation expansion in the vertex coupling of
a Hermitian matrix model free energy of the general form \cite{bipz}
\begin{equation}
F = {1 \over M^2 } \log \int \prod_i  D \Phi_i \exp ( - S )
\label{free}
\end{equation}
in the limit $M \rightarrow \infty$, where the $\Phi_i$ are sufficient
$M \times M$ Hermitian matrix variables to describe the matter decoration and the $\log$
ensures that only connected graphs contribute.
For the Ising model on trivalent planar
graphs, for instance, we need two matrices $X,Y$ and the action takes
the form
\begin{equation}
S = Tr \left\{\frac{1}{2} ( X^2 + Y^2) - g X Y - \frac{\lambda}{3} \left[ e^h X^3 +
e^{-h} Y^3 \right],
\right\}
\label{ising}
\end{equation}
where $g = \exp ( - 2 \beta )$ and $h$ is the external field. We can see that 
the $X^3$ terms can be thought of as spin-up or $+$ vertices and the $Y^3$ terms spin-down
or $-$, while the inverse of the quadratic terms gives the appropriate edge weights
for $+ \; +$, $+ \;  -$ and $- \;  -$ edges. 

To date there has been less consideration of arrow vertex models
on ensembles of planar random graphs and none at all of bond vertex models.
In the arrow case one decorates the edges of the graph with arrows and assigns
weights to the vertices depending on the local arrow configurations
around each vertex, in
the bond case one simply has empty or occupied edges (bonds) and vertex weights
depending on the local bond configuration.
Some  
arrow vertex models have been discussed using elementary means
in \cite{gins, des1}, mostly on 4-valent random graphs, revealing
$XY$ and Ising critical behaviour for suitable coupling constant loci.
More recently some elegant and sophisticated work using character expansions
\cite{ABAB} has provided a full solution of an arrow vertex model on 4-valent
planar graphs confirming the $XY$ critical point found in \cite{gins}.

In this paper we give the first solutions of  {\it bond} vertex models
on planar random graphs. 
Our main example is a 4-vertex model
on $\Phi^3$ random graphs, but the general method works for any bond 
vertex models whose weights depend only on the number
of bonds at a vertex rather than their particular cyclic ordering. On regular
lattices such models are termed symmetric vertex models and we retain
the nomenclature here.
Since our 4-vertex model on $\Phi^3$ random graphs
is the random graph equivalent of the symmetric 
8-vertex model on the honeycomb lattice discussed 
by Wu et.al. \cite{Wu1, Wu2, Wu3} we take our cue from 
\cite{Wu1} where the honeycomb lattice model was first solved
by making use
of the so-called
generalised weak graph transformation \cite{wegnag} between vertex weights. 
This brought the partition function to a form that was recognisable as that of an Ising model
in field an the critical behaviour of the model was then deduced from the corresponding Ising results.

In what follows 
we first 
define our 4-vertex model, then solve it by mapping onto the Ising model
in field on $\Phi^3$ planar random graphs, which itself was solved in \cite{bk}.
We move on to discuss higher valency symmetric models and show that
the same general method of solution still works.
We then compare these results
with those 
of \cite{Wu1,Wu2,Wu3} for the 8-vertex model
on the honeycomb lattice
and note that the generalized weak graph symmetry of the honeycomb lattice model
finds a counterpart in an orthogonal variable transformation in the matrix integrals
used to define the random graph models.
Finally, we close 
with some speculations on the relations of vertex models to
Lorentzian metric quantum gravity, where they can be thought of as providing
a discretisation of causal structure.

\section{The 4-Vertex Model on 3-valent random graphs}

The canonical (fixed number of vertex) partition function
for the 4-vertex model on $\Phi^3$ planar random graphs
is given by
\begin{equation}
Z_{4v} = Z_{4v} ( a, b, c, d) = \sum_{\Phi_N} \sum_{G}  a^{N_1} {b^{N_2}} c^{N_3} 
d^{N_4}
\label{parth}
\end{equation}
where there are two  summations: $\sum_{\Phi_N}$ 
over $\Phi^3$ random graphs with $N$ vertices
and $\sum_{G}$
over all possible bond configurations on each $\Phi^3$ graph
built using the vertices
of Fig.1.
Each graph has  $N_1$ vertices of type $(a)$,
$N_2$ of type $(b)$, $N_3$ of type $(c)$, $N_4$ of type $(d)$
and $N = N_1 + N_2 + N_3 + N_4$.

We can generate the graphs we require, suitably decorated with the allowed bond
configurations, by inserting the action 
\begin{eqnarray}
S = Tr \left\{ \frac{1}{2} ( X^2 + Y^2) -  \lambda \left[{ a \over 3}
X^3 +  { d \over 3}  Y^3
+ c  X Y^2 +   b X^2 Y \right] \right\}.
\label{efat}
\end{eqnarray}
into equ.(\ref{free}), where
$X,Y$ are $M \times M$ Hermitian matrices
and we have introduced an overall vertex coupling $\lambda$ for convenience..
The limit $M \rightarrow \infty$ picks out the planar diagrams.
A  direct attempt at solving a matrix model with the action equ.(\ref{efat}) might be possible
using the character expansion techniques of \cite{ABAB}, but it
turns out to be easier
to follow in the footsteps of \cite{Wu1,Wu2} and establish a correspondence with an Ising model
in order to determine the critical behaviour of the vertex model.

We start by 
by carrying out
the orthogonal transformations
\begin{eqnarray}
X &\rightarrow& ( X + Y ) / \sqrt{2} \nonumber \\
Y &\rightarrow& (X - Y) / \sqrt{2}
\label{ortho0}
\end{eqnarray}
on the matrices in the Ising action
of equ.(\ref{ising}), followed by the rescalings $X \rightarrow X / ( 1 - g )^{1/2}, \; Y
\rightarrow Y / ( 1 + g)^{1/2}$, $\lambda \rightarrow \sqrt{2} \lambda  (
1 - g)^{3/2}$. This change of variable
in the matrix integral has a trivial Jacobian and gives the new action
\begin{equation}
S = Tr \left\{\frac{1}{2} ( X^2 + Y^2) -  {\lambda \cosh(h) \over 3 } \left[ X^3 +3 g^* X Y^2 \right] -
{\lambda \sinh ( h ) (g^*)^{3/2} \over 3}  \left[  Y^3 + { 3 \over g^*} X^2 Y \right] \right\}.
\label{ibond0}
\end{equation}
where $g^* = ( 1 - g) / ( 1 + g)$, which is clearly of the same general form as
equ.(\ref{efat}). The bond vertex weights from equ.(\ref{ibond0}) are also
given in Fig.1.
It should be remarked that the transformation of equ.(\ref{ortho0})
is the duality transformation for the Ising model on random $\Phi^3$ graphs \cite{Ising}, and that the
bond graphs generated by the transformed action in equ.(\ref{ibond0}) are precisely
the high-temperature expansion graphs of the original Ising model
on $\Phi^3$ graphs.

We can see that the vertex weights coming from the Ising model in equ.(\ref{ibond0}), 
which we denote by tilde'd quantities,
\begin{eqnarray}
\tilde a &=& \cosh (h) \nonumber \\
\tilde b &=& \sinh (h) (g^*)^{1/2} \nonumber \\
\tilde c &=& \cosh (h) g^* \nonumber \\
\tilde d &=& \sinh (h )  ( g^* )^{3/2}
\end{eqnarray}
are fixed in the ratio $\tilde a \tilde d = \tilde b \tilde c$, which will
{\it not} be the case for the generic weights of equ.(\ref{efat}). 
However, if we consider the effect of the further orthogonal rotation
\footnote{This preserves the propagator $X^2 + Y^2$ in the action of equ.(\ref{ibond0})
and again has a trivial Jacobian, so it does not effect the evaluation of the matrix integral.}
\begin{equation}
\left( \begin{array}{c}  X \\ Y \end{array} \right) \rightarrow  \left( \begin{array}{cc}
                                   \cos (\theta) &   \sin ( \theta ) \\
                                  -\sin ( \theta )  & \cos (\theta )  \\
                                   \end{array} \right) \left( \begin{array}{c} X \\ Y \end{array} \right)
\label{XYrot}
\end{equation}
on the vertex weights of equ.(\ref{efat}), we find
\begin{eqnarray}
\tilde a &=& {1 \over ( 1 + y^2)^{3 /2 } } \left[ a + 3 y b + 3 y^2 c + y^3 d \right] \nonumber \\
\tilde b &=& {1 \over ( 1 + y^2)^{3 /2 } } \left[ - y a + (1 - 2 y^2 ) b - ( y^3 - 2 y ) c + y^2 d \right] \nonumber \\
\tilde c &=& {1 \over ( 1 + y^2)^{3 /2 } } \left[ y^2 a + ( y^3 - 2 y ) b
+ ( 1 - 2 y^2 ) c + y d \right] \label{weak3} \\
\tilde d &=& {1 \over ( 1 + y^2)^{3 /2 } } \left[ - y^3 a + 3 y^2 b - 3 y c
+ d \right] \nonumber 
\end{eqnarray}
where we have extracted $\cos (\theta)$, denoted $\tan ( \theta )$ as $y$
and deliberately used tilde'd quantities again for the transformed weights.
We can now demand that these transformed weights satisfy
$\tilde a \tilde d = \tilde b \tilde c$, 
just as in equ.(\ref{ibond0}) which has the effect of making the partition function of the vertex model equivalent to
that of the Ising model in field.
Setting $\tilde a \tilde d = \tilde b \tilde c$ gives
the following equation for the rotation parameter $y$
from equating the right hand sides of equ.(9)
\begin{equation}
B y^2 + 2 ( C - A ) y - B = 0
\label{yeq}
\end{equation}
where $A = b d - c^2$, $B = a d - b c$, $C = a c - b^2$. Since $ - \infty \le y
\le \infty$
is at our disposal we can solve this for general $a,b,c,d$
to obtain transformed weights satisfying $\tilde a \tilde d = \tilde b \tilde c$.

To deduce the critical behaviour of the vertex model by using this correspondence with the 
Ising model in field we need the relation between the vertex weights and the Ising
parameters $\beta$ and $h$.
The Ising and tilde'd vertex model parameters were related by
\begin{eqnarray}
\tanh ( \beta ) &=& { \tilde c \over \tilde a} \nonumber \\
\tanh ( h ) &=& { \tilde b \over \sqrt{ \tilde a \tilde c } }.
\label{iv}
\end{eqnarray}        
as can be seen directly from equ.(\ref{ibond0}).
We can then substitute for the tilde'd weights in terms of the  
original weights and $y$ using equ.( 9 ).
to give the relation between the original weights $a,b,c,d$ and 
the Ising parameters $\beta, h$
\begin{eqnarray}
\exp ( 2 \beta ) &=& { B y + C - A \over  A + C  } \nonumber \\
\tanh ( h ) &=& {W \over T } \left( { B y + 2 C \over B y - 2 A } \right)^{1/2},
\label{params}
\end{eqnarray}
where $y$ is again the one of the solutions of equ.(\ref{yeq}),
$A,B,C$ are as above, $T= (b + d ) y + a + c$ and $W= ( b + d ) 
- ( a + c ) y$.       

On $\Phi^3$ random graphs the Ising model displays a third order magnetisation transition
in zero field.
The second of equs.(\ref{params}) shows that 
zero Ising field implies that $W=0$, which in terms of the vertex weights
(when $y$ is substituted for) gives the locus
\begin{equation}
a ( b^3 + d^3 ) - d ( a^3 + c^3 ) + 3 (a b + b c + cd ) 
(c^2 - b d - b^2 + a c ) = 0.
\label{Vzero}
\end{equation}
From the vertex model
perspective it is most natural to disallow loops of length one and two in the random graphs
in order to make sure that there are no double edges or tadpoles,
in which case the Ising critical temperature is given by $\beta_c = { 1 \over 2} \log \left( {103 \over 28} \right)
= 0.7733185$ \cite{bk2}. The first of equs.(\ref{params}) then gives
\begin{equation} 
{103 \over 28} = { \sqrt{ ( a c - b^2 - b d + c^2)^2 + ( a d - b c )^2}  \over ac - b^2 + b d - c^2 } 
\label{betac}
\end{equation}
as the position of the third order critical point on the locus equ.(\ref{Vzero}).
We still have, of course, the echo of the field driven Ising transition appearing for 
$\beta> \beta_c$ along the locus equ.(\ref{Vzero}).

\section{Higher valency vertex models}

In this section we show that
on 4-valent ($\Phi^4$) random graphs
the Ising equivalency can still 
be easily established by rotating the variables 
in a matrix model action. One extra ingredient in the 
4-valent case compared with the 3-valent case is that
the cyclic order of the bonds allows one to distinguish
between two different sorts of vertices with two occupied and two empty edges,
one with occupied bonds at right angles and one with 
occupied bonds ``straight through''. 
The Ising equivalence only holds for the symmetric case in which equal
weights are assigned to these bond configurations. 

On $\Phi^4$ random graphs the Ising model action is
\begin{equation}
S = Tr \left\{\frac{1}{2} ( X^2 + Y^2) - g X Y + \frac{\lambda}{4} \left[
e^h X^4 +
e^{-h} Y^4 \right],
\right\}
\label{ising4}
\end{equation}
with only a higher order potential distinguishing the action from
that for $\Phi^3$ graphs in equ.(\ref{ising}). The change of variables
$X \rightarrow ( X + Y ) / \sqrt{2}$, $Y \rightarrow (X - Y) / \sqrt{2}$
and rescalings $X \rightarrow X / ( 1 - g
)^{1/2}, \; Y
\rightarrow Y / ( 1 + g)^{1/2}$, $\lambda \rightarrow 2 \lambda (1 -g )^2$ gives the action
\begin{eqnarray}
S &=& Tr \left\{
\frac{1}{2} (X^2 + Y^2) + {\lambda \cosh (h) \over 4 } \left[ X^4 + (g^*)^2 Y^4 \right]
+ { \lambda \cosh (h ) g^* \over 2 } \left[ 2 X^2 Y^2 + X Y X Y \right] \right. \nonumber \\
&+& \left. { \lambda \sinh ( h) g^* } \left[ \sqrt{g^*} X Y^3 + { 1 \over \sqrt{g^*}} X^3 Y \right] 
\right\}
\label{v4}
\end{eqnarray}
where, as on $\Phi^3$ graphs, $g^* = ( 1 - g ) / ( 1 + g)$. 
Note the presence of the $\left[ 2 X^2 Y^2 + X Y X Y \right]$ term
fixing the weights of right angled and straight through two-bond vertices
to be the same.
It should be remarked once again that
the change of variables in equ.(\ref{ortho0}) is a duality transformation for the model, and
that equ.(\ref{v4}) generates the high temperature expansion graphs of the original Ising model
in equ.(\ref{ising4}).

The effect of the orthogonal rotation
of equ.(\ref{XYrot}) on the vertices arranged in some suitable lexicographic order
(e.g. $X^4$, $X^3 Y$, $X^2 Y^2$, $X Y^3$, $Y^4$
as labelled on Fig.2) gives 
\begin{eqnarray}
\tilde a &=& {1 \over ( 1 + y^2)^{2 } } \left[ a + 4 y b + 6 y^2 c + 4 y^3 d + y^4 e \right] \nonumber \\
\tilde b &=& {1 \over ( 1 + y^2)^{2 } } \left[ - y a + (1 - 3 y^2 ) b - ( 3 y^3 - 3 y ) c + ( 3 y^2 - y^4 ) d + y^3 e \right] \nonumber \\
\tilde c &=& {1 \over ( 1 + y^2)^{2 } } \left[ y^2 a + ( 2 y^3 - 2 y ) b + ( y^4  - 4 y^2 + 1) c + (2 y - 2 y^3) d + y^2 e \right] \\
\tilde d &=& {1 \over ( 1 + y^2)^{2 } } \left[ - y^3 a + ( 3 y^2  - y^4  ) b - (3 y - 3 y^3) c + (1 - 3 y^2 ) d + y e \right] \nonumber \\
\tilde e &=& {1 \over (1 + y^2)^2}      \left[ y^4 a - 4 y^3 b + 6 y^2 c - 4 y d + e \right].
\label{weak4}
\end{eqnarray}  

The strategy for solving the $\Phi^4$ model is identical to the $\Phi^3$ case. One uses the 
above transformations of the vertex weights
to bring the model to the Ising action defined above in equ.(\ref{v4}). 
The tilde'd weights 
in this ($\tilde a$ for $X^4$ etc.) can be characterised in various ways,
for example: $ \tilde a \tilde d = \tilde d \tilde c$, $\tilde c^2 = \tilde a \tilde e$. Having achieved this the critical 
behaviour can again be read off from that of the Ising model in field, this time on $\Phi^4$ graphs.
Indeed, it is clear that generic symmetric vertex models on random graphs 
can be solved by using this method. 
In the q-valent case one carries out the same  sequence of two orthogonal rotations
on the action for the Ising model on $\Phi^q$ planar random graphs
\begin{equation}
S = Tr \left\{\frac{1}{2} ( X^2 + Y^2) - g X Y + \frac{\lambda}{q} \left[
e^h X^q +
e^{-h} Y^q \right]
\right\}.
\label{isingN}
\end{equation}
to obtain the action for the symmetric vertex model. 
One can thus deduce that the critical behaviour of the
symmetric vertex model on q-valent random graphs will be Ising-like.

\section{Comparison with the Symmetric 8-Vertex Model on the honeycomb lattice} 

The closest regular lattice equivalent of our $\Phi^3$ planar random graphs
is the honeycomb lattice where each vertex still has coordination number three
but, in contrast to the random graphs, all loops are of length six.
The eight possible bond vertex configurations all look similar to those for
the random lattice in Fig.1 but we now have orientational order
as well as cyclic order around each vertex so $b$ and $c$ represent
three possible orientations each. However if we impose equality
between the different orientations, we arrive at the symmetric vertex
model of \cite{Wu1,Wu2,Wu3} which, like the $\Phi^3$ random graph vertex model,
has 4 distinct vertex weights.
The partition function is given in a similar fashion to equ.(\ref{parth})
by 
\begin{equation}
Z_{8v} = Z_{8v} ( a, b, c, d) = \sum_{G}  a^{N_1} {b^{N_2}} c^{N_3} {d^{N_4}}
\label{parthh}
\end{equation}
where there is now no sum over random graphs since all the bond 
configurations are living on the honeycomb lattice.                     

Several symmetry properties of the partition function are immediately
apparent. From the negation of vertices
occurring in pairs we have $Z_{8v}(a,b,c,d) = Z_{8v}(-a,-b,-c,-d)=Z_{8v}(-a,b,-c,d)=Z_{8v}(a,-b,c,-d)$.
Similarly, exchanging dark and light bonds gives
$Z_{8v}(a,b,c,d)=Z_{8v}(d,c,b,a)$. Perhaps rather less obvious is the generalised weak graph
symmetry \cite{wegnag} $Z_{8v} ( a, b , c , d) = Z_{8v}(\tilde a ,\tilde  b ,\tilde  c ,\tilde  d )$,
where 
the transformations between the original and tilde'd variables are precisely
those generated by the orthogonal rotations in the
random graph model in equ.(\ref{weak3}).

In \cite{Wu2} the generalised weak graph
symmetry on the honeycomb lattice was 
thought of as being generated by rotations with
\begin{equation}
V(y) = { 1 \over (1 + y^2)^{1 / 2}} \left( \begin{array}{cc}
                                   1 & y \\
                                   y & -1 \\
                                   \end{array} \right)
\label{wg1}
\end{equation}
on the vertices in ``bond space'', where occupied and empty bonds on a given edge 
were to be thought of as
components of a vector. An alternative choice of transformation
\begin{equation}
U(y) = { 1 \over (1 + y^2)^{1 / 2}} \left( \begin{array}{cc}
                                   1 & y \\
                                   -y & 1 \\
                                   \end{array} \right)
\label{wg2}
\end{equation}
was related to $V(y)$ by the negations $\tilde b \rightarrow - \tilde b$, 
$\tilde d \rightarrow - \tilde d$
and was hence not independent. Remarkably, the second transformation
is just that generated
by equ.(\ref{XYrot}) in the random graph model.

The generalised weak graph symmetry played an important role in the
original solution of the honeycomb lattice
model  \cite{Wu2} because
a suitable choice of $y$ allowed generic vertex weights $a,b,c,d$ to be 
transformed
to $\tilde a,\tilde b,\tilde c,\tilde d$ satisfying $\tilde a \tilde d =\tilde
 b \tilde c$
just as for the random lattice.
For vertex weights satisfying the latter
condition
\begin{equation}
Z_{8v} (\tilde a,\tilde b,\tilde c,\tilde d ) = ( \tilde a \cosh (h) / 2)^N ( \
cosh (\beta ) )^{- 3 N / 2} Z_{Ising} ( \beta , h)\label{highT}
\end{equation}
where $Z_{Ising}$ was the standard honeycomb lattice Ising partition function.
The equivalence followed from observing
that this choice of weights meant that the vertices generated precisely the
right weights for the
diagrams of the high temperature
expansion of the Ising model on the honeycomb lattice. 
                      
The critical behaviour of the symmetric 8-vertex model on the honeycomb lattice
was thus deduced by following exactly the same path we have taken for the 
4-vertex model on $\Phi^3$ random graphs -- exploiting a symmetry
of the partition function to transform it to that of an Ising model in field.
The parallels run even closer since equs.(\ref{weak3}, \ref{yeq},
\ref{iv}, \ref{params}, \ref{Vzero})
still hold identically for the honeycomb lattice vertex model and the
only difference in equ.(\ref{betac}) is in the numerical value of
the Ising critical temperature on the left hand side.
This might grandiloquently be phrased as a non-renormalization theorem
for the vertex weights since the ratios of weights in equ.(\ref{iv})
are unaffected by the coupling to gravity that is represented by the sum
over random graphs.

The close correspondence between the $\Phi^3$ and honeycomb results
becomes a little less surprising when one considers a second
determination of the critical behaviour of the honeycomb lattice model
by Wu in \cite{Wu3}
where it was shown that a direct mapping between the Ising and vertex
models was possible when a decoration-iteration transformation was employed.
In Fig.3 we show a single vertex, which
hosts a spin subject to an external field $H$. In addition
the edges have spins which are subject to a field
$2 H'$ and the two sorts of spins interact via a spin-spin coupling $R$.
The presence of a dark bond
can be denoted by an edge spin $\sigma=1$ and its absence by $\sigma = - 1$.
The vertex model weights can then be
represented correctly in terms of appropriate $H',H$ and $R$
\begin{eqnarray}
\cosh ( 2 R ) &=& { B \over 2 ( A C )^{1/2} } \nonumber \\
\exp ( 4 H' ) &=& { C \over A} \nonumber \\
\cosh ( 2 H ) &=& { 2 b c \over \sqrt{ A C } } \left( {B^2 \over 4 A C } - { B \over 4 b c } - 1 \right)
\end{eqnarray}
when the central vertex spins
are traced over. One can go in the other direction and decimate the edge Ising spins
by replacing the two $R$ interactions and edge field $2H'$
with a single interaction $\beta$ and fields at the end of each edge     
(i.e. on the vertices).
This gives one a standard honeycomb lattice Ising model in field.
The relation between the vertex model weights and the Ising parameters
$\beta, h$ determined in this manner
is identical to that equ.(\ref{params}).

This decoration-iteration approach strongly supports the 
observation here that the equivalence
between the symmetric 8-vertex model on the honeycomb
lattice and the Ising model in field
continues to hold for the 4-vertex model 
on  $\Phi^3$ random graphs.
The decoration iteration transformation is a local transformation
which relies only on the valency of each vertex
and in both cases all vertices have valency three. 
It also
supports the result that the relation between the Ising
and vertex parameters is identical on
$\Phi^3$ random graphs and the honeycomb lattice, since the
decoration-iteration transformation doesn't ``see'' the randomness,
so as far as it is concerned the random and honeycomb lattices are identical 
\footnote{One possible fly in the ointment is that
loops of length one and two, i.e. double edges (``bubbles'')
and self-loops (``tadpoles''), are in principle present in the random graphs.
The star-triangle relation would degenerate at vertices adjacent to such edges.
The simplest solution,
which we have taken, is to restrict the ensemble of random graphs to exclude such
configurations, which does not affect the critical behaviour of the Ising model
though non-universal quantities such as the critical temperature may be changed.}.                                                     

The critical behaviour of symmetric vertex models 
on higher valency regular lattices may be deduced by using the appropriate
generalized weak graph transformations to map the models onto Ising models 
in field in all cases. The
generalised weak graph transformation written down by Wu for general valency $q$
\begin{equation}
W_{ij} = {1 \over ( 1 + y^2)^{q/2}} \sum_{k=0}^j
\left( \begin{array}{c} { i} \\ {k } \end{array} \right)
\left( \begin{array}{c} { q - i } \\ {  j - k } \end{array} \right)  ( -1)^k y^{i + j - 2 k}
\label{wooo}
\end{equation}
where the vertices are labelled in some suitable order as in the $\Phi^{3}, \;
\Phi^{4}$ cases, is precisely that
obtained by picking out the appropriate powers in an expansion of terms such as
$(\cos ( \theta) X + \sin (\theta ) Y )^{q-k} ( -\sin ( \theta) X + \cos ( \theta ) Y )^k $.

\section{Speculations on Connections with Lorentzian Gravity}

$\Phi^3$ graphs are duals to triangulations which are the natural
discretisation of {\it Euclidean} signature $(+ \; +)$ 2D spacetimes.
An analytical continuation to physical Lorentzian $( - \; + )$ signatures
is by no means on such firm ground as the analogous procedure
in standard quantum field theory on flat backgrounds.
It is therefore of some interest to attempt to formulate
discretised theories which might serve as toy models for Lorentzian
gravity. 

An essential extra ingredient compared with
the Euclidean case for any Lorentzian theory
is a causal structure. 
One then faces the option of whether
causal structures should be summed over or
imposed. A relatively rigid choice was explored
in \cite{janrenate} where triangulations with
spacelike slices joined by fluctuating timelike
edges were used to triangulate Lorentzian
space. It was found that only when branchings
were allowed was the critical behaviour of Euclidean
gravity recovered upon analytical continuation
back.

Another possibility is to postulate that global,
or at least long-range, causal structure
might emerge dynamically in much the same way that
the fractal structure of the 2D manifolds does in the Euclidean case.
Inspired by \cite{MS,M}
we introduce local causal structure 
by using spacelike edges to triangulate our Lorentzian spacetime.
The normals to all 
the triangles are then time-like, which can be
either past or future directed. It is then natural to return
to the dual spin network picture, which gives an
embedded {\it directed} $\Phi^3$ graph if we
denote the causal relations
between points in the interior of adjacent triangles by arrows
pointing from the past to the future of each vertex.
This should be contrasted with Euclidean 2D triangulations where
the dual picture is of undirected $\Phi^3$ graphs.         

It is thus natural to use arrow vertex models 
in this context, 
rather than the bond vertex models considered
so far. However the translation between
arrows and bonds turns out to be trivial, as we see below. 
The various
sorts of  3-valent ($\Phi^3$) arrow vertices are most 
conveniently labelled by their (indegree, outdegree).
The different possibilities are $(3,0)$, $(0,3)$,
$(2,1)$ and $(1,2)$
as shown in Fig.4.  A natural choice is to take conjugate weights
for the $(3,0)$ and $(0,3)$ vertices and similarly for the
$(2,1)$ and $(1,2)$ vertices as this
preserves the symmetry under arrow reversal. The  partition function
for this 4-arrow-vertex model is thus of the form
\begin{equation}
Z = \sum_{\Phi_N} \sum_{G} a^{N_1} {\bar a^{N_2}} b^{N_3} {\bar b^{N_4}}
\label{part}
\end{equation}   
where the first sum is again over different planar 
$\Phi^3$ graphs, the second over arrow
assignments and we have
$N_1$ $(3,0)$ vertices, $N_2$ $(0,3)$ vertices etc.
Note that only the $(2,1)$ and $(1,2)$ vertices can be considered as
regular spacetime points since the $(3,0)$ and $(0,3)$ vertices act as
microscopic black and white holes respectively.

We can obtain this partition function from the $N$ vertex term
in the expansion of the free energy of the
complex matrix model with the action
\begin{eqnarray}
S = \frac{1}{2} Tr \left\{ \Phi^{\dagger} \Phi - { \alpha \over 3} \left[(\Phi^{\dagger})^3 + \Phi^3 \right]
- i { \beta \over 3} \left[(\Phi^3 - (\Phi^{\dagger})^3 \right]
-  \gamma \left[(\Phi^{\dagger})^2 \Phi + \Phi^2 (\Phi^{\dagger}) \right]
- i \delta \left[(\Phi^{\dagger})^2 \Phi - \Phi^2 (\Phi^{\dagger}) \right] \right\}
\label{arrows}
\end{eqnarray}
where $\Phi$ is now an $M \times M$ {\it complex} matrix
and $a = \alpha + i \beta, \; b = \gamma + i \delta$.                                                      
We can transform simply to a bond vertex formulation of the model
by splitting $\Phi$
into Hermitian components  $X + i Y$.
This allows us to interpret $Y$ edges as those
containing bonds and $X$ edges as empty
(or {\it vice-versa}).
The resulting action in terms of $X,Y$
\begin{eqnarray}
S = Tr \left\{ \frac{1}{2} ( X^2 + Y^2) -  { (\alpha + 3 \gamma) \over 3} X^3 -  { (\beta + 3 \delta) \over 3}  Y^3
- ( \gamma -  \alpha ) X Y^2 -  ( \delta -  \beta )X^2 Y \right\}.
\label{erealfat}
\end{eqnarray}
is now clearly that of our 4-bond-vertex model on $\Phi^3$ random graphs.         
This displays Ising criticality, as we have seen, so a suitable
tuning of the vertex couplings in equ.(\ref{erealfat}) (and hence in the
original arrow vertex model of equ.(\ref{arrows}) ) can reach this point
\footnote{Interestingly, the presence of the black hole and white hole vertices
appears to be necessary to do this, there is insufficient freedom with only
the regular vertices.}.
Since this is a continuous transition one might hope that the diverging
correlations of the Ising model at the critical point might be translated 
back to the appearance of long range casual 
structure ex nihilo in the arrow vertex
model formulation. However, 
one can see heuristically that closed loops of arrows are present at all scales
at the transition point. Cluster algorithms exist for vertex models
\cite{cluster} which act by identifying such closed loops and flipping them. These are effective at combatting
critical slowing down so they are identifying and flipping loops at all
scales near a continuous transition point. 
The loops are innocuous from the 
statistical mechanical point of view, but they represent closed timelike 
loops when the vertex arrows are interpreted as giving the causal structure.

A profusion of  closed timelike loops is obviously an undesirable
property
and we thus conclude that the particular 
vertex model discussed here is probably not
a suitable candidate for modeling discretised Lorentzian gravity
as it stands. It would be interesting to determine what modifications might 
make it so. 

\section{Discussion}

We have solved a bond 4-vertex model on an ensemble
of planar $\Phi^3$ graphs by taking our lead from the
honeycomb lattice solution of the symmetric 8-vertex
model and exploiting
its equivalence to an Ising model in field.
An important ingredient of the solution was an orthogonal
rotation in the matrices use to define the random graph model in order
bring a generic weight configuration on to the Ising locus. This turned
out to be functionally identical to the generalised weak graph transformation used in the original
honeycomb lattice solution \cite{Wu1}. The method of solution and the equivalence of the matrix rotation
to the generalised weak graph transformation on an equivalent regular lattice were also shown to work for
higher valency symmetric vertex models on random graphs. 

We have seen that the relations between the vertex model parameters and the Ising parameters were
identical for the honeycomb and random graph models. 
This is not so surprising at it might first seem when
viewed in the light of the 
decoration-iteration solution on the honeycomb lattice, since this depends
only on the valency of the vertices, which is identical in the random
case (modulo caveats about
tadpoles and bubbles discussed previously). 
This result could be couched as a non-renormalization theorem for the
appropriate ratios of vertex weights, since putting the models on ensembles
of random graphs is equivalent to coupling them to 2D quantum gravity.

The methods discussed in this paper are restricted to symmetric vertex models since
the orthogonal rotation of equ.(\ref{ortho0})
necessarily gives only symmetric weights 
when applied to Ising actions such as those in equs.(\ref{ising},
\ref{ising4},\ref{isingN}). 
From the matrix model point of view the vertex model solution shows that
an apparently hopeless potential containing $X^3, Y^3, X^2 Y$ and $X Y^2$ terms
can still give rise to a soluble model, once the matrices are appropriately
transformed. Finally, we have speculated on the relation between 
vertex models and Lorentzian signature gravity, pointing out a
potential problem with closed timelike loops when
employing the class of models considered here.

\section{Acknowledgements}

This work was partially supported by a Leverhulme Trust Research Fellowship
and a Royal Society of Edinburgh/SOEID Support Research Fellowship.

%

%
\clearpage \newpage
\begin{figure}[htb]
\vskip 20.0truecm
\includegraphics{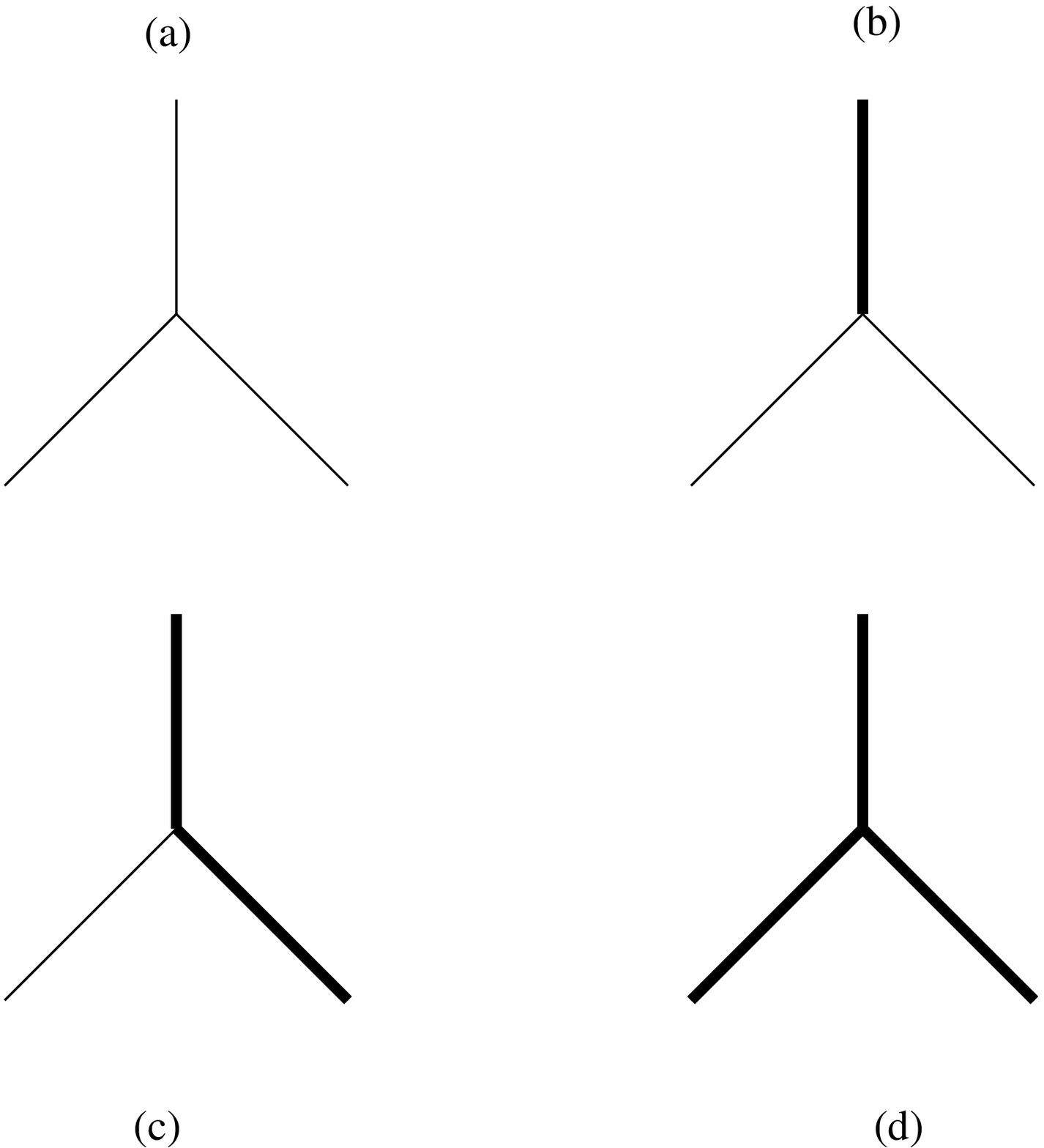}
\caption[]{\label{fig1} The possible bond vertices
which appear in the model, using the same
notation as in \cite{Wu1}. The Ising weights on the random lattice, which can be read off from equ.(\ref{ibond0}), are:
$\tilde  a  \; = \cosh (h), \; \tilde b \; = \sinh (h) (g^*)^{1/2}$, 
$\tilde  c \; =  \cosh (h) g^* $ and $\tilde d \; =  \sinh (h )  ( g^* )^{3/2} $.}
\end{figure}
\clearpage \newpage
\begin{figure}[htb]
\vskip 20.0truecm
\includegraphics{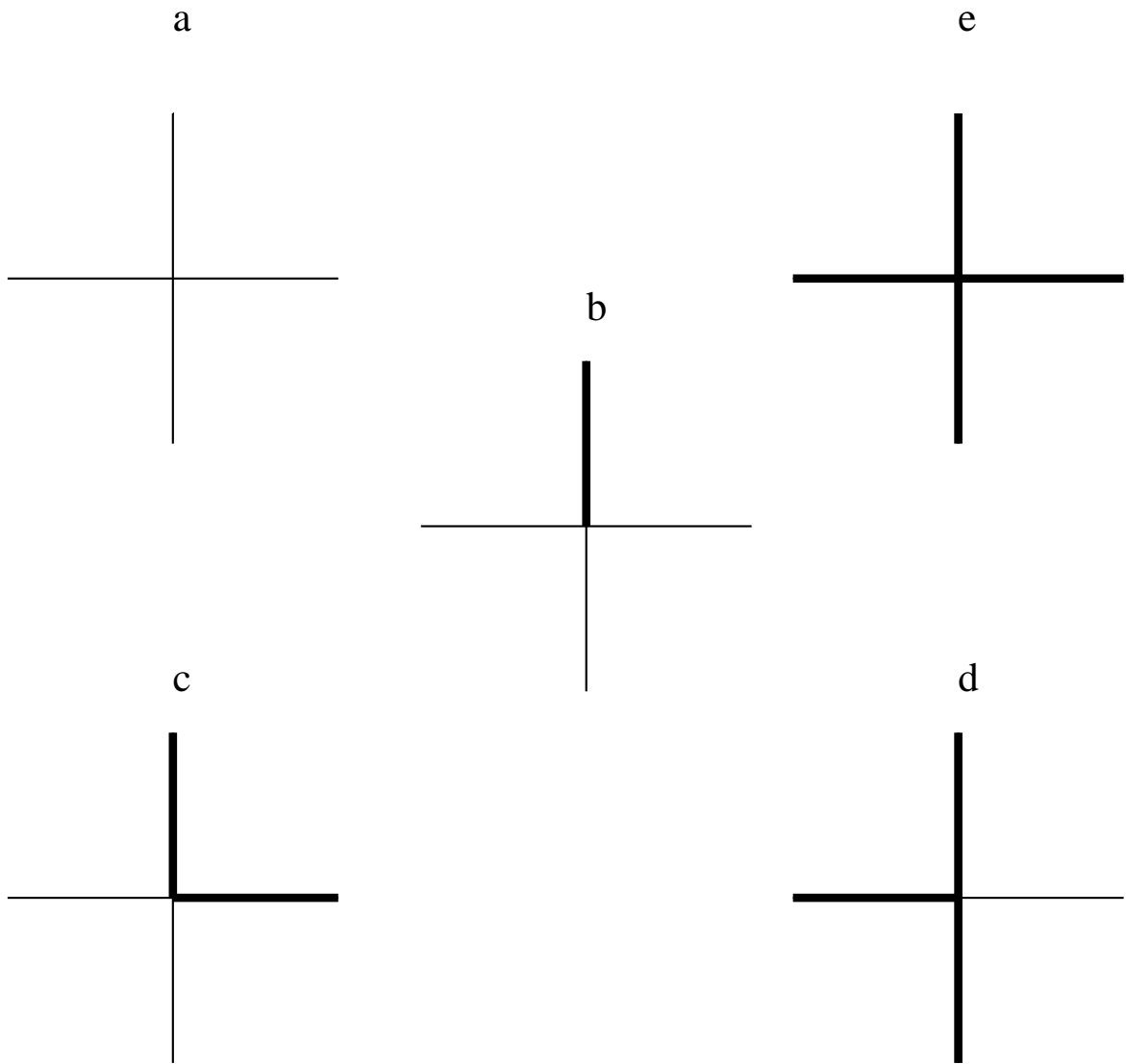}
\caption[]{\label{fig4} The 5 different vertex weights in the
  symmetric 16 vertex model on the square lattice.
  The ``straight-through'' two bond configuration which is not shown also receives
  weight $c$.
 }
\end{figure}                             
\clearpage \newpage
\begin{figure}[htb] 
\vskip 20.0truecm
\includegraphics{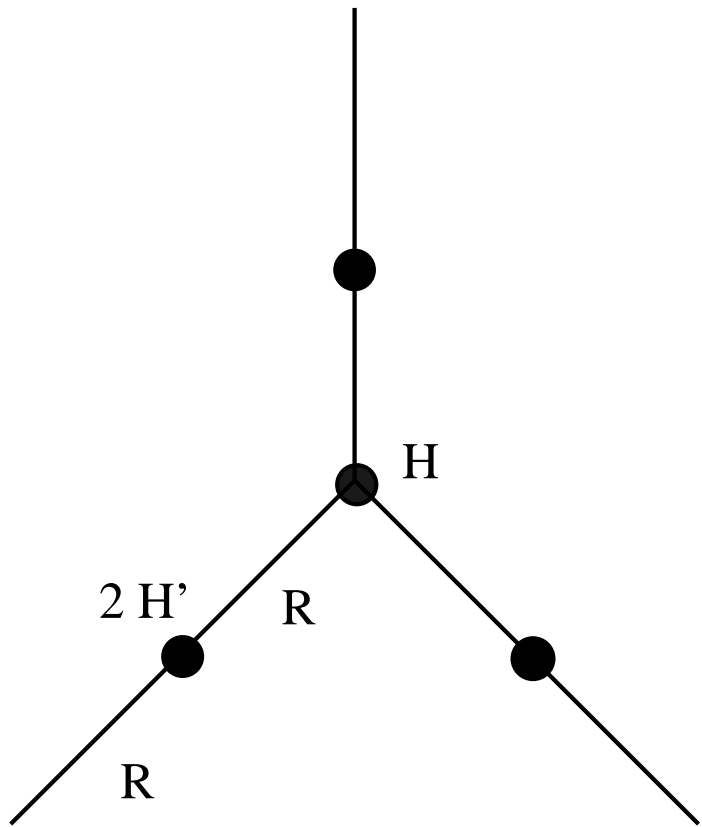}
\caption[]{\label{fig3} The decorating spins 
on the edges are subject to an
external field $2 H'$, the central spin to a field $H$.
The edge-vertex spin interactions have weight $R$.}
\end{figure}
\clearpage \newpage
\begin{figure}[htb]
\vskip 20.0truecm
\includegraphics{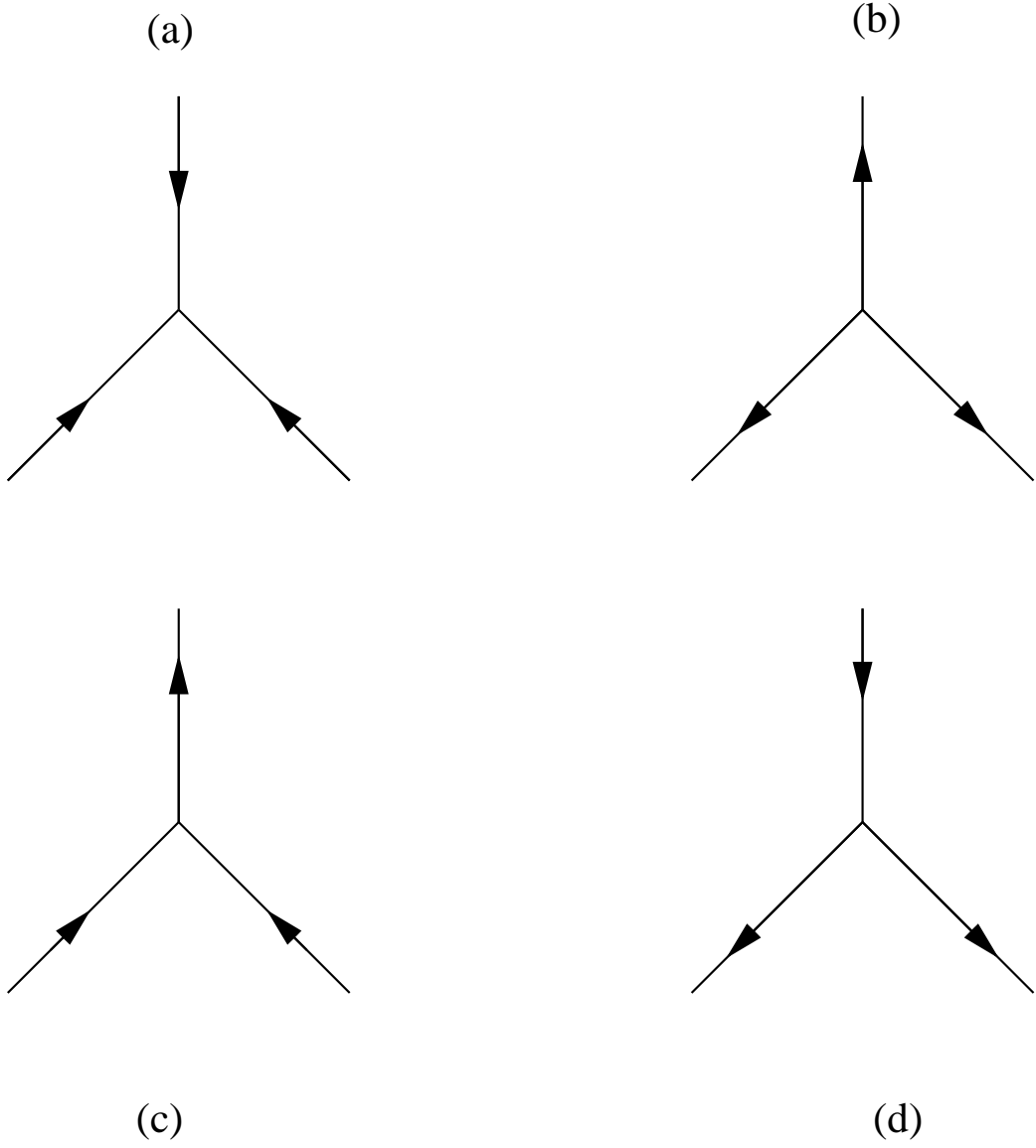}
\caption[]{\label{fig2} The possible arrow vertices
which appear in the model. Their weights
are equivalent to linear combinations of the bond vertex weights
in Fig.1. On the random lattices
the corresponding terms in the action are :(a)  $(\Phi^{\dagger})^3$, (b) $\Phi^3$, (c) $(\Phi^{\dagger})^2 \Phi$, (d) $\Phi^2 \Phi^{\dagger}$}
\end{figure}
\end{document}